\documentclass[10pt]{iopart}
\usepackage{graphicx,ifthen,iopams,setstack,amssymb}
\jl{1}
\usepackage[latin1]{inputenc}
\newcommand{\writer}{gerhard}

\eqnobysec
\begin{document}
\title{Interaction and thermodynamics of spinons in the XX chain} 
\author{
Michael Karbach$^{1,2}$, 
Gerhard M{\"u}ller$^2$,
and Klaus Wiele$^{1,2}$
}
\address{
$^1$Fachbereich Physik, Bergische Universit{\"a}t Wuppertal, 
  42097 Wuppertal, Germany \\
  $^2$Department of Physics,
  University of Rhode Island,
  Kingston RI 02881, USA \\
}

\ifthenelse{\equal{\writer}{gerhard}}%
{\date{\today~--~1.10}} 
{\date{\version}}
\begin{abstract}
  The mapping between the fermion and spinon compositions of eigenstates
  in the one-dimensional spin-1/2 $XX$ model on a lattice with $N$ sites is
  used to describe the spinon interaction from two different perspectives: (i)
  For finite $N$ the energy of all eigenstates is expressed as a function of
  spinon momenta and spinon spins, which, in turn, are solutions of a set of
  Bethe ansatz equations. The latter are the basis of an exact thermodynamic
  analysis in the spinon representation of the $XX$ model. (ii) For $N\to\infty$ the
  energy per site of spinon configurations involving any number of spinon
  orbitals is expressed as a function of reduced variables representing
  momentum, filling, and magnetization of each orbital. The spins of spinons in
  a single orbital are found to be coupled in a manner well described by an
  Ising-like equivalent-neighbor interaction, switching from ferromagnetic to
  antiferromagnetic as the filling exceeds a critical level.  Comparisons are
  made with results for the Haldane-Shastry model.
\end{abstract}
%
\section{Introduction}\label{sec:intro}
%

To the few and precious situations, where interacting quantum many-body systems
can be analyzed by rigorous methods in great detail, belong a host of models for
quantum spin chains. Prominent among them is the $XX$
model for exchange-coupled electron spins on a one-dimensional lattice,
\begin{equation}
  \label{eq:17}
 \mathcal{H}_{XX}\doteq\sum_{\ell=1}^N[S_\ell^xS_{\ell+1}^x+ S_\ell^yS_{\ell+1}^y].
\end{equation}
Periodic boundary conditions are assumed throughout. The $XX$ model is the
special case $\Delta=0$ of the $XXZ$ model. The latter also has terms $\Delta
S_\ell^zS_{\ell+1}^z$ in its Hamiltonan. The $XXZ$ model is exactly solvable via
Bethe ansatz for arbitrary values of $\Delta$ \cite{CG66, YY66abc}.

The composition of all $XX$ eigenstates by free, spinless Jordan-Wigner
fermions has been the basis of most advances in the study of the
thermodynamics, correlation functions, and dynamics \cite{LSM61, Kats62,
  Niem67, KHS70, MBA71, BJ76, CP77, VT78, MS84a, MPS83ab, IIKS93, SNM95,
  DKS00}. The alternative composition of the same states provides a framework
for the interpretation of dominant features observed in the spectra and
intensity distributions of dynamical quantities. Spinons have spin 1/2 and
semionic exclusion statistics \cite{Hald91a}. The one-to-one mapping between
the fermion and spinon compositions was presented in a recent study of the
$m$-spinon dynamic structure factors based on product formulas for transition
rates \cite{AKMW06}. Here we use the fermion-spinon mapping for different
purposes.

In preparation of our main themes -- interaction and thermodynamics of spinons
-- we introduce alternative quantum numbers for fermion momenta and for spinon
momenta and spins, along with rules that translate the fermion composition of
any $XX$ eigenstate into the corresponding spinon composition
(Sec.~\ref{sec:qpr}). For the energy of an arbitrary $XX$ eigenstate, the
mapping converts its dependence on the fermion quantum numbers into its
dependence on the spinon quantum numbers. The resulting expression is akin to a
coordinate Bethe ansatz (CBA) for the spinon momenta and spins
(Sec.~\ref{sec:spint}).  From an entirely different perspective, the $XX$ chain
is interpreted as a set of interacting spinon orbitals with internal degrees of
freedom exhibiting features akin to electrons in partially filled electronic
shells (Sec.~\ref{sec:inspor}).  A thermodynamic Bethe ansatz (TBA) for spinons
is constructed from the CBA discussed in Sec.~\ref{sec:spint} and then analyzed
exactly (Sec.~\ref{sec:tdsp}). Several aspects of spinon interaction and spinon
thermodynamics discussed for the $XX$ model are compared with corresponding
properties of the Haldane-Shastry (HS) model (\ref{sec:appb}).


%
\section{$XX$ spectrum from top down and 
bottom up}\label{sec:qpr}
%
The complete spectrum of eigenstates of the $XX$ model is described via
complementary sets of quasiparticles with different exclusion statistics and
with their vacua at opposite ends of the energy range: (i) the free
Jordan-Wigner lattice fermions and the closely related partially interacting
magnon solutions of the CBA, (ii) the interacting spinons.
Only the fermions are free. Therefore, it is desirable to perform calculations
in the fermion representation. On the other hand, it is desirable to use the
spinons for the interpretation of the spectrum and the dynamics because the
ground state (physical vacuum) coincides (for even $N$) with the spinon vacuum.
The magnons are important because they are natural products of the
CBA, the method on which most advances in the analysis of the $XXZ$ model rest
\cite{Taka99}. 

%
\subsection{Magnons and fermions}\label{sec:magnons}
%
The structure of the Bethe ansatz equations (BAE) for the states with
magnetization $M_z=N/2-r$ of the $XXZ$ model,
\begin{equation}
  \label{eq:7}
  e^{\imath Nk_{i}} = \prod_{j\neq i}^r\left[
  -\frac{1+e^{\imath(k_{i}+k_{j})}-2\Delta e^{\imath
      k_{i}}}{1+e^{\imath(k_{i}+k_{j})}-2\Delta e^{\imath k_{j}}}\right], 
  \quad i=1,\ldots,r,
\end{equation}
undergoes a drastic simplification in the $XX$ limit $\Delta\to0$:
\begin{equation}
  \label{eq:8}
   e^{\imath Nk_i} = (-1)^{r-1},\quad i=1,\ldots,r,
\end{equation}
with solutions
\begin{equation}
  \label{eq:9}
   k_i=\pi-\frac{2\pi}{N}I_i,
\end{equation}
provided $k_i+k_j\neq\pi$ holds for all pairs of magnon momenta.
The Bethe quantum numbers $I_i$ are integers for odd $r$ and half-integers for even
$r$ over the range $|I_i|\leq N/2$. This simple rule becomes more complicated for
$\Delta\neq0$ \cite{Taka99}.

In those instances, where magnon pairs with momenta $k_i+k_j\to\pi$ exist, the
limit $\Delta\to0$ in (\ref{eq:7}) is singular, and nontrivial solutions exist
\cite{BKMW04}. The solution of the BAE as a limit process then yields both real
and complex-conjugate solutions. The critical magnon pairs do not contribute to
the energy of the eigenstate,
\begin{equation}
  \label{eq:10}
  E= \sum_{i=1}^r \cos k_i.
\end{equation}
Thus removing one critical pair from an eigenstate yields a degenerate
eigenstate with $\Delta k=\pi$ and $\Delta M_z=2$. This kind of degeneracy is
predicted and quantitatively described by the $sl_2$ loop algebra symmetry for
the case $\Delta=0$ \cite{DFM01,FM01ab}. Critical pairs occur only for even
$N$. If $N$ is odd the criticality condition for the associated Bethe quantum
numbers, $I_i+I_j=N/2~ \mathrm{mod}~N$, is impossible to satisfy, given that
$I_1,I_2$ are either both integers or both half-integers.

The $XX$ Hamiltonian (\ref{eq:17}) can be mapped onto a system of free,
spinless lattice fermions by means of the Jordan-Wigner transformation followed
by the Fourier transform \cite{LSM61}:
\begin{equation}
  \label{eq:12}
\mathcal{H}_{XX} = \sum_{\{p_i\}}\cos p_i\,c_{p_i}^\dagger c_{p_i}^{},
\end{equation}
where the sum is over all sets of distinct fermion momenta from the allowed values
\begin{equation}
  \label{eq:13}
 p_i=\frac{\pi}{N}\,\bar{m}_i,\quad
\bar{m}_i\in \left\{ 
\begin{tabular}{ll}
${\displaystyle \{1,3,\ldots,2N-1\}}$ & for even $N_f$ \\
${\displaystyle \{0,2,\ldots,2N-2\}}$ & for odd $N_f$
\end{tabular} \right.. 
\end{equation}
The number of fermions in an eigenstate depends on the magnetization $M_z$:
\begin{equation}
  \label{eq:15}
  N_f=\frac{N}{2}-M_z,\qquad -\frac{N}{2}\leq M_z\leq\frac{N}{2}.
\end{equation}
For even $N$ the ground state of $\mathcal{H}_{XX}$ is unique. It has
$M_z=0$ and thus contains $N_f=N/2$ fermions. For odd $N$ the ground state
is fourfold degenerate. 

How are the fermions and the magnons, which originate from different methods of
analysis, related to one another? It is evident from (\ref{eq:9}) and
(\ref{eq:13}) that the non-critical magnons have momenta and energies that
exactly correspond to fermion momenta and energies.  Whereas the fermions are
free, the non-critical magnons are not. They scatter off each other
elastically. However, the associated phase shift is $\pi$ for all such
events. This reflects the well-known equivalence between hard-core bosons and
free fermions in one dimension.

Regarding eigenstates with critical magnon pairs we note that they are at least
twofold degenerate within their invariant Hilbert subspace of given quantum
numbers $k$ and $M_z$. This degeneracy between states with $\Delta k=0$ and
$\Delta M_z=0$ suggests that the magnon eigenbasis for states with
critical pairs is related to the fermion eigenbasis by an orthogonal
transformations within the invariant $(k,M_z)$-subspaces. The critical and
non-critical magnons of the $XX$ model can be interpreted as fragments of
string excitations in the context of the CBA applied to the $XXZ$ model. The
limit $\Delta\to0$ is singular in the BAE (\ref{eq:7}) but the singularities
can be removed by a basis transformation.

One consequence is that the thermodynamics of magnons as analyzed via TBA for
the $XXZ$ model becomes equivalent, for $\Delta\to0$, to the thermodynamics of
free fermions \cite{Taka99}. Another consequence is that the advances recently
reported in the calculation of transition rates via algebraic Bethe ansatz for
the $XXZ$ model \cite{BKMW04} can be translated into corresponding advances,
for $\Delta\to0$, in the fermion representation of the $XX$ model
\cite{AKMW06}.


%
\subsection{Spinons}\label{sec:fermspin}
%
The unique ground state of $\mathcal{H}_{XX}$ for even $N$ coincides with the
vacuum for spinon quasiparticles. The total number of spinons is confined
to the range $0\leq N_s\leq N$ and restricted to be even (odd) if $N$ is even
(odd). Knowledge of the numbers of spinons with spin up and spin down
determines both $N_s$ and $M_{z}$.
\begin{equation}
  \label{eq:lj24}
 N_+ +N_-=N_s,\qquad N_+ -N_-= 2M_z.
\end{equation}

The range of momentum values for spinon orbitals was first determined in the
context of the HS model \cite{Hald91a}. Minor adaptations are necessary for the
$XX$ model \cite{AKMW06}.  The orbitals available for occupation by spinons are
equally spaced at $\Delta\kappa=2\pi/N$ and their number is $(N-N_s)/2+1$, where
$N_s=0,2,\ldots,N$ for even $N$ and $N_s=1,3,\ldots,N$ for odd $N$.  The allowed
orbital momentum values are\footnote{Note the slight change in convention
  from Ref. \cite{AKMW06} regarding the $m_i$ for odd $N$.}
\begin{equation}
  \label{eq:lj66}
  \kappa_i=\frac{\pi}{N}\,m_i,\quad m_i = \frac{N_s}{2}, \frac{N_s}{2}+2, \ldots,
  N-\frac{N_s}{2}.  
\end{equation}
Each available orbital may be occupied by spinons of either spin orientation
without further restrictions. The spinon quantum numbers $m_i^\sigma$ are thus
integers for even $N$ and half-integers for odd $N$.  The $XX$ eigenstate
specified by spinon quantum numbers $\{m_{i}^{\sigma}\}$ has wave number
\begin{equation}
  \label{eq:lj67}
  k= \left(\frac{\pi}{N}\sum_{\sigma=\pm} 
\sum_{j_\sigma=1}^{N_\sigma}m_{j_\sigma}^\sigma -\frac{N\pi}{2}\right)~ 
\mathrm{mod}(2\pi).
\end{equation}

\begin{figure}[b]
\centering
\includegraphics[width=61mm]{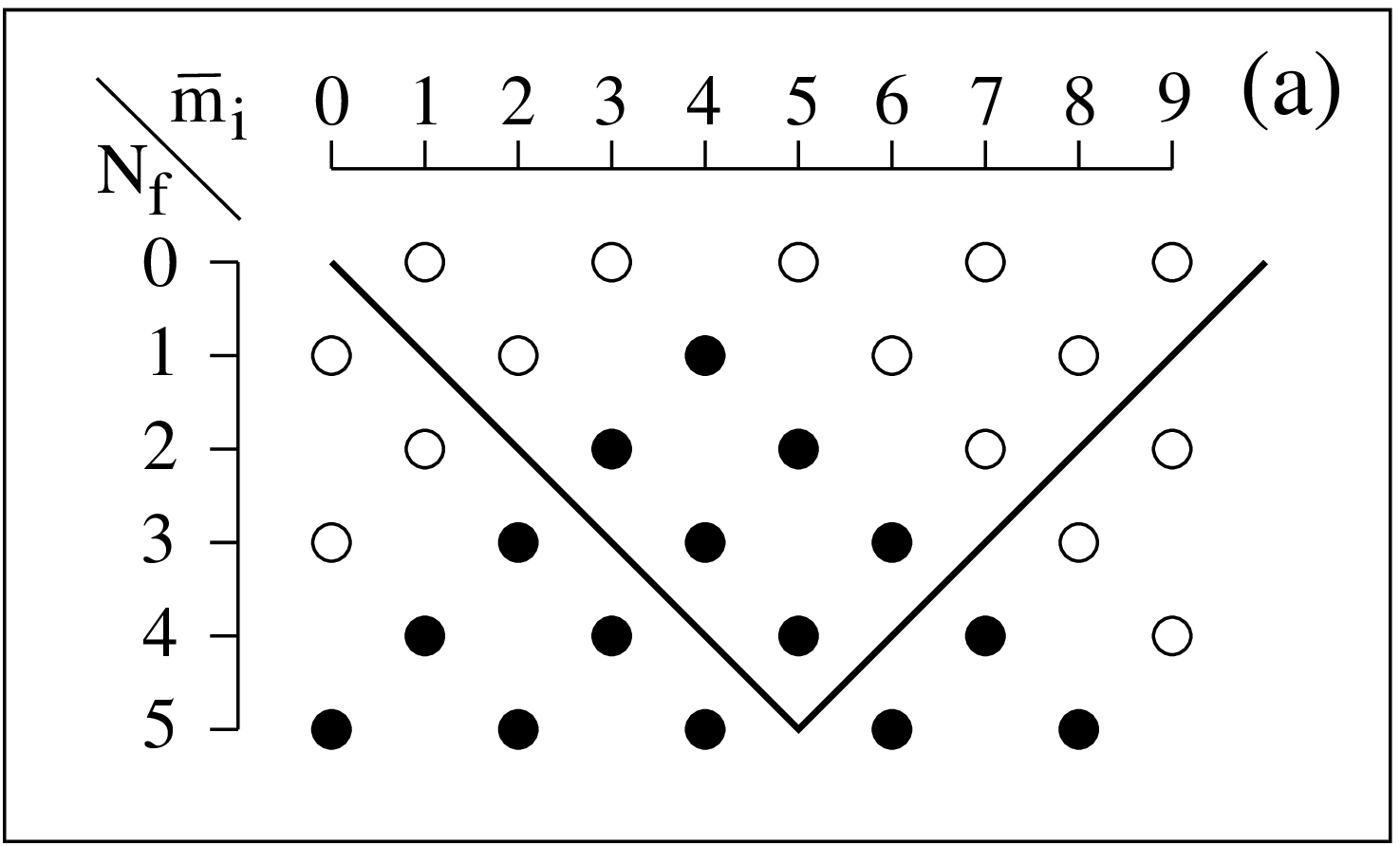}
\hspace*{5mm}%
\includegraphics[width=53.7mm]{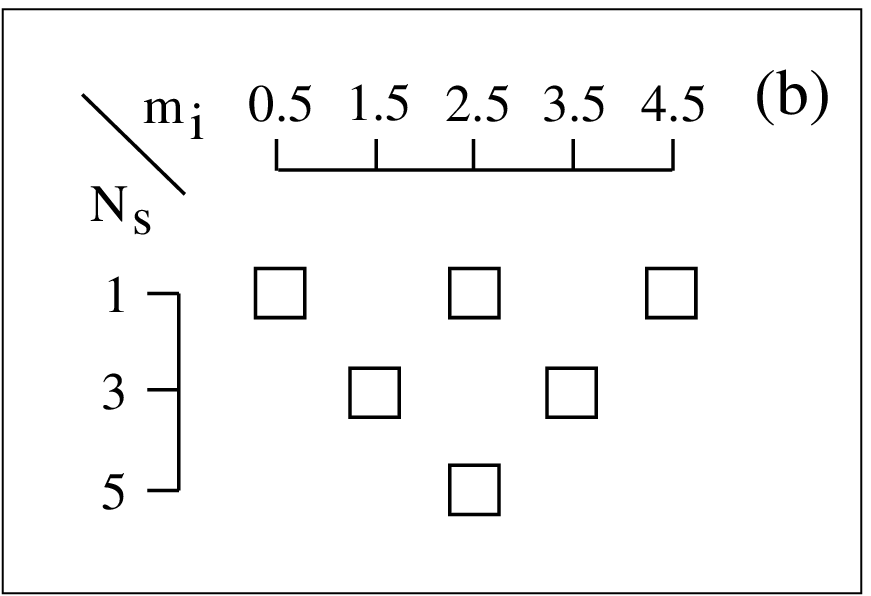}

\caption{(a) Fermion orbitals available to $N_f$ fermions and (b) spinon
  orbitals available to $N_s$ spinons in $XX$ eigenstates for $N=5$. Fermion
  orbitals can be either vacant (open circle) or singly occupied (full circle).
  The particular fermion configuration shown in each row represents one lowest
  energy states for given $N_f$. Spinon orbitals can be either vacant or
  occupied by up to $N_s$ spinons with arbitrary spin polarization. No specific
  spinon configuration is shown.}
  \label{fig:fskN5}
\end{figure}

\begin{figure}[b]
  \centering
  \includegraphics[width=110mm]{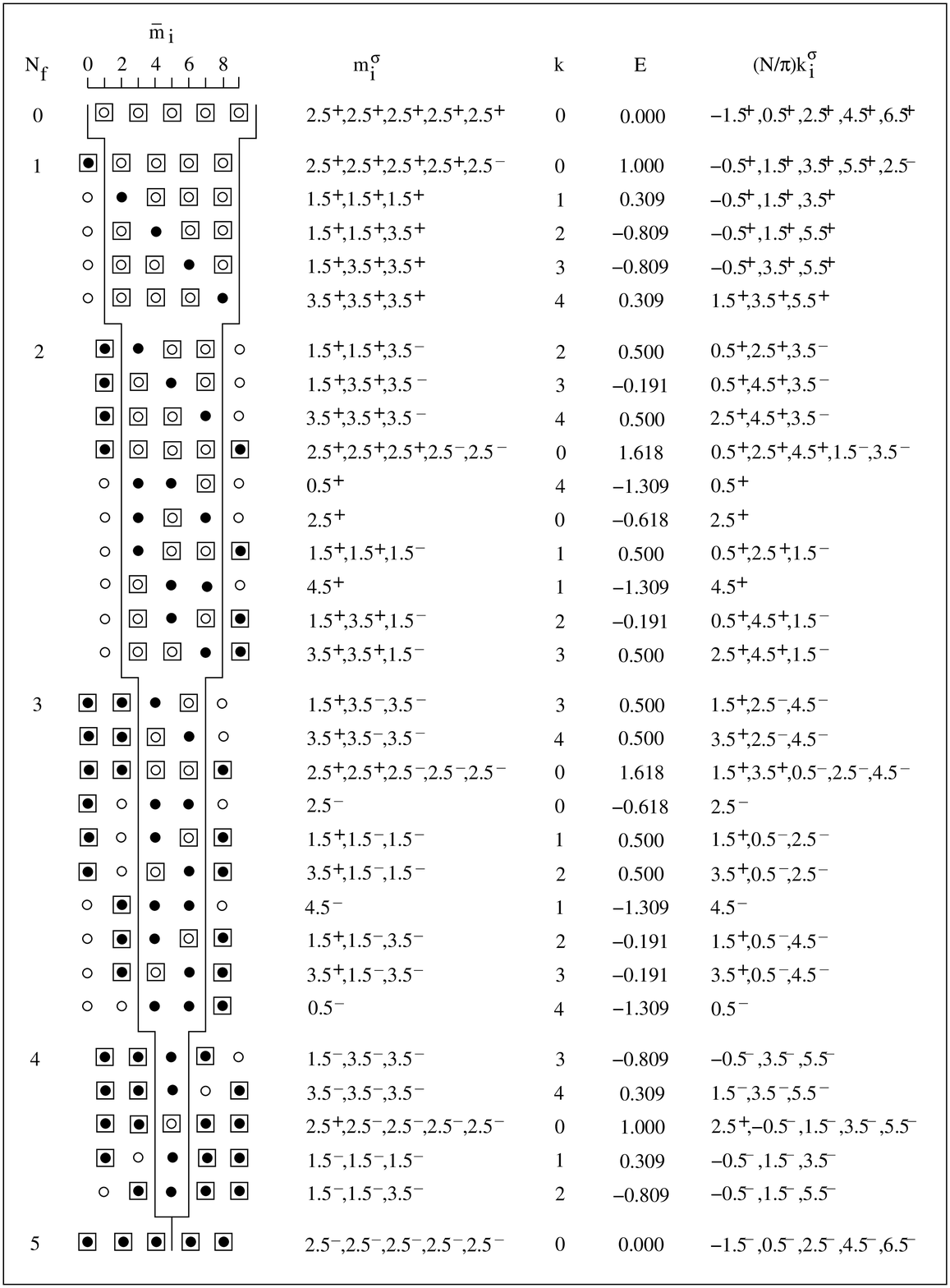}  
  \caption{Fermion configurations of all eigenstates for $N=5$. Fermionic
    particles (holes) are denoted by full (open) circles.  Spinon particles
    with spin up (down) are denoted by squares around open (full) circles. The
    fermion quantum numbers $\bar{m}_i$ can be read off the diagram. The spinon
    quantum numbers $m_i^\sigma$ are given explicitly and can be inferred from the
    fermion configuration as explained in the text. Also given are the wave
    number $k$ (in units of $2\pi/N$) and the energy $E$ of each $XX$
    eigenstate. The spinon momenta $k_i^\sigma$ are discussed in
    Sec.~\ref{sec:basp}.}
  \label{fig:fermspinN5}
\end{figure}

In Ref. \cite{AKMW06} we demonstrated how to keep track of the spinons in the
fermion representation for a system with $N=4$, where the (unique) ground state
is the spinon vacuum. Here we consider $N=5$ for illustration, where the ground
state consists of four 1-spinon states. The set of allowed fermion momentum
states for $N_f=0,1,\ldots,5$ is shown in Fig.~\ref{fig:fskN5}(a) and the sets of
allowed spinon momentum states for $N_s=1,3,5$ in Fig.~\ref{fig:fskN5}(b).  An
expanded version of Fig.~\ref{fig:fskN5}(a) is shown in
Fig.~\ref{fig:fermspinN5} with all $2^N=32$ distinct fermion configurations
(circles) grouped according to $N_f$ and with the associated
spinon configurations (squares) indicated. The $\lor$-shaped line of
Fig.~\ref{fig:fskN5}(a) becomes the forked line in Fig.~\ref{fig:fermspinN5}.

The exact spinon configuration is encoded in the fermion configuration as
described in the following: (i) Consider the $\lor$ or the fork as dividing the
fermion momentum space into two domains, the inside and the outside. The
outside domain wraps around at the extremes ($\bar{m}_i=N$ mod$(N)=0$).  (ii)
Every fermionic hole (open circle) inside represents a spin-up spinon (square
surrounding open circle) and every fermionic particle (full circle) outside
represents a spin-down spinon (square surrounding full circle).  (iii) Any
number of adjacent spinons in the representation of Fig.~\ref{fig:fermspinN5}
are in the same orbital of Fig.~\ref{fig:fskN5}(b).  Two spin-up (spin-down)
spinons that are separated by $\ell$ consecutive fermionic particles (holes)
have quantum numbers separated by $2\ell$.  (iv) The spinon quantum numbers are
sorted in increasing order from the right-hand prong of the fork toward the
left in the inside domain $(m_i^+)$ and toward the right with wrap-around
through the outside domain $(m_i^-)$.


%
\section{Interacting spinons}\label{sec:spint}
%
Naturally, the key to studying the spinon interaction for the $XX$ model is the
mapping between the fermion composition and the spinon composition of every
eigenstate, combined with the fact that the fermions are free. Here we study
this interaction on the level of spinon {\em particles}, then, in
Sec.~\ref{sec:inspor}, on the level of spinon {\em orbitals}.

%
\subsection{Energy of spinon configurations}\label{sec:genexp}
%
Consider an arbitrary eigenstate of $\mathcal{H}_{XX}$ for finite (even or odd)
$N$ in the fermion eigenbasis. Given the configuration of fermion quantum numbers
$\{\bar{m}_1,\ldots,\bar{m}_{N_f}\}$ of the state selected, the mapping described in
Sec.~\ref{sec:fermspin} produces a unique configuration of spinon quantum numbers
with range restricted as in (\ref{eq:lj66}):
\begin{equation}
  \label{eq:48}
  \frac{N_s}{2}\leq m_1^\sigma\leq m_2^\sigma\leq\cdots
\leq m_{N_\sigma}^\sigma\leq N-\frac{N_s}{2},\quad \sigma=\pm.
\end{equation}
Note that we have sorted them into two groups according to spin orientation
and, within each group, in ascending order. The number of distinct
$m_i^\sigma$-configurations satisfying (\ref{eq:48}) for fixed $N$ and $N_s$ is
$(N+1)!/[N_s!(N+1-N_s)!]$. Summation over $N_s$ yields $2^N$. All eigenstates
are accounted for. Their energies are found to have the following dependence on
the spinon quantum numbers:
\begin{equation}
  \label{eq:26}
E\big(\{m_{j_+}^+\},\{m_{j_-}^-\}\big)= E_0(N_+,N_-)+ 
\sum_{\sigma=\pm}\, \sum_{j_\sigma=1}^{N_\sigma}
\sin\big(\kappa_{j_\sigma}^\sigma-\tau_{j_\sigma}^\sigma\big)
\end{equation}
with
\begin{equation}
  \label{eq:lj33}
  \kappa_{j_\sigma}^\sigma =\frac{\pi}{N}\,m_{j_\sigma}^\sigma,\quad 
  \tau_{j_\sigma}^\sigma= \frac{\pi}{N}\big(N_\sigma+1-2j_\sigma\big),
\quad \sigma=\pm,
\end{equation}
and where the reference energy
\begin{equation}
  \label{eq:18}
  E_0(N_+,N_-)= \sum_{\{\bar{m}_i^0\}}\cos\left(\frac{\pi}{N}\,\bar{m}_i^0\right),
\end{equation}
\begin{eqnarray}
  \label{eq:19}
  \{\bar{m}_i^0\} &=& \left\{\frac{1}{2}(N+N_+-N_-+2), 
\frac{1}{2}(N+N_+-N_-+6),\ldots
  \right. \nonumber \\
  && \hspace*{35mm} \left. \ldots,\frac{1}{2}(3N-N_++N_--2)\right\},
\end{eqnarray}
only depends on $N_+ -N_-$, i.e. on $M_z$. For even $N$, $E_0$ is the energy of
the lowest eigenstate for given $M_z$, but for odd $N$ it does not represent
the energy of any eigenstate.

The spinons are not free, notwithstanding the fact that expression
(\ref{eq:26}) is a sum of 1-spinon terms. The spinon interaction is hidden in
the sorting criterion (\ref{eq:48}).  Moving one spinon into a different
orbital will, in general, affect more than just one term in (\ref{eq:26}), and
switching the spin of one spinon affects all terms including the reference
energy.
  
%
\subsection{Bethe ansatz equations for 
spinons}\label{sec:basp}
%
The very structure of Eq.~(\ref{eq:26}) exhibits features characteristic of a
CBA solution for spinons undergoing two-body elastic scattering. When we employ
the spinon quantum numbers (\ref{eq:48}) in the role of Bethe quantum numbers
(BQN) for the spinon configuration, Eq.~(\ref{eq:26}) can be rewritten in the
form
\begin{equation}
  \label{eq:49}
  E\big(\{m_{j_+}^+\},\{m_{j_-}^-\}\big)= E_0(N_+,N_-)+ 
\sum_{\sigma=\pm}\, \sum_{j_\sigma=1}^{N_\sigma}
\sin k_{j_\sigma}^\sigma
\end{equation}
with a universal energy-momentum relation $\epsilon(k)=\sin k$ for spinons,
provided the spinon momenta $k_{i}^\sigma$ satisfy the BAE
\begin{equation}
  \label{eq:50}\hspace*{-5mm}
  Nk_i^\sigma= \pi m_i^\sigma +\sum_{\sigma'=\pm}
\sum_{j=1}^{N_{\sigma'}}\theta_{XX}(k_i^\sigma-k_j^{\sigma'}),\quad
  i=1,\ldots,N_\sigma,\quad \sigma=\pm,
\end{equation}
\begin{equation}
  \label{eq:46}\hspace*{-5mm}
  \theta_{XX}(k_i^\sigma-k_j^{\sigma'})
=\pi\,\mathrm{sgn}(k_i^\sigma-k_j^{\sigma'})\delta_{\sigma\sigma'}.
\end{equation}
Only spinons with parallel spins scatter off each other. All spinon momenta
$k_i^\sigma$ of given spin orientation are distinct:
\begin{equation}
  \label{eq:1}
  k_{j_\sigma}^\sigma=
\frac{\pi}{N}\left(m_{j_\sigma}^\sigma-N_\sigma-1+2j_\sigma\right),\quad
  j_\sigma=1,\ldots,N_\sigma,\quad \sigma=\pm.
\end{equation}
They are sorted in increasing order and bounded as follows:
\begin{equation}
  \label{eq:2}\hspace*{-19mm}
  \frac{\pi}{N}\left(\frac{1}{2}N_s-N_\sigma+1\right)\leq
  k_1^\sigma<k_2^\sigma<\cdots<k_{N_\sigma}^\sigma\leq 
\frac{\pi}{N}\left(N-\frac{1}{2}N_s+N_\sigma-1\right).
\end{equation}
The distance of any $k_i^\sigma$ from the upper or lower bound is
$\ell(2\pi/N)$, where $\ell=0,1,\ldots,(N-N_s)/2+N_\sigma-1$. The complete sets
of $m_i^\sigma$ and $k_i^\sigma$ for $N=5$ are shown in
Fig.~\ref{fig:fermspinN5}.

Even though all spinon momenta of given spin orientation are distinct from one
another, their exclusion statistics is semionic, not fermionic. This is
demonstrated by applying the defining relation \cite{Hald91a}
\begin{equation}
  \label{eq:3}
  \Delta d_\sigma=-\sum_{\sigma'=\pm}g_{\sigma\sigma'}\Delta N_{\sigma'}
\end{equation}
for the statistical interaction coefficients $g_{\sigma\sigma'}$ to the
situation described by (\ref{eq:2}), taking into account that the number of
available momentum states $\Delta d_\sigma$ with $N_s=N_+ +N_-$ spinons already
present is affected both by the next particle added and by the shifting bounds.
The result is $g_{\sigma\sigma'}=1/2$ for all combinations of spinon spin
orientations. With all these spinon properties in the $XX$ model established we
are ready to analyze their thermodynamics via TBA from the bottom up. This is
the theme of Sec.~\ref{sec:tdsp}.

%
\section{Interacting spinon orbitals}\label{sec:inspor}
%
Meanwhile we look at the $XX$ chain from an entirely different perspective.
Instead of considering individual spinons moving along the chain and scattering
off each other as described by the BAE (\ref{eq:50}) we focus on spinon
orbitals, specified by orbital momenta $\kappa_i$, and populate them with
spinons.

Each spinon orbital in isolation acquires an energy that varies systematically
as spinons with spin up or down are added, producing effects not unlike those
familiar from electronic shells in atomic physics. If two or more spinon
orbitals are populated with spinons we can express the total energy as a sum of
intra-orbital and inter-orbital terms.
  
%
\subsection{Spinons in one orbital}\label{sec:spin1orb}
%
There are $N+1$ states with $N$ spinons. According to (\ref{eq:lj66}), all $N$
spinons are then in the orbital with momentum $\kappa=\pi/2$. These $N+1$
states all have the same wave number, $k=0$, but different values of
magnetization, $M_z=-N/2, -N/2+1,\ldots,N/2$.\footnote{In the context of the
  $XXZ$ model at $\Delta=1$ all these states are degenerate, forming the
  multiplet with total spin $S_T=N/2$.}

The state with all spinon spins up is the fermion vacuum.  Flipping one spinon
spin at a time translates into adding two fermions to the orbitals with the
highest available energies in the band. The energy levels thus follow from
recursive sequences, one for even $N_f$ and one for odd $N_f$:
  \begin{equation}
    \label{eq:20}\hspace*{-10mm}
    E(N_f)=E(N_f-2)+2\cos\frac{(N_f-1)\pi}{N},\quad E(0)=0,\quad E(1)=1.
  \end{equation}
  The fermionic particle-hole symmetry implies $E(N-N_f)=E(N_f)$.  The level
  sequence resulting from (\ref{eq:20}) is characterized by spacings ranging
  between $\Delta E\sim\mathrm{O}(1)$ at the bottom $(N_f\ll N)$ and $\Delta
  E\sim\mathrm{O}(N^{-1})$ at the top $(N_f\lesssim N/2)$. In the limit
  $N\to\infty$ we convert Eq.~(\ref{eq:20}) into an integral that yields the
  reduced energy $\epsilon\doteq E/N$ as a function of the reduced
  magnetization $m_z\doteq M_z/N$:
  \begin{equation}
    \label{eq:21}
    \epsilon= \frac{1}{\pi} \cos(\pi m_z),\quad 
-\frac{1}{2}\leq m_z\leq +\frac{1}{2}.
  \end{equation}

  The level distribution implied by (\ref{eq:21}) can be explained
  qualitatively by a simple microscopic model. The spinon-spin coupling within
  the given spinon orbital is well represented by a ferromagnetic
  equivalent-neighbor Ising interaction:
  \begin{equation}
    \label{eq:22}
    \mathcal{H}_e=-J_e\sum_{i<j}\left[\sigma_i\sigma_j-1\right],\quad \sigma_i=\pm1.
  \end{equation}
  The energy-level spectrum of $\mathcal{H}_e$ is
\begin{equation}
  \label{eq:23}
  E_e=\frac{1}{2}J_e\left[N^2-(2M_z)^2\right],\quad M_z=\frac{N}{2}-N_f.
\end{equation}
If we set $J_e=2/ \pi N$ we obtain from (\ref{eq:23}) the following functional
dependence for the reduced energy $\epsilon_e\doteq E_e/N$ on the reduced
magnetization $m_z$:
\begin{equation}
  \label{eq:27}
  \epsilon_e=\frac{1}{\pi}\left(1-4m_z^2\right),\quad  
-\frac{1}{2}\leq m_z\leq +\frac{1}{2}.
\end{equation}
It shares with (\ref{eq:21}) several properties: (i) identical values at $m_z=0$
and $m_z=\pm1/2$, (ii) a quadratic dependence at $|m_z|\ll1/2$, (iii)
a linear dependence at $m_z\simeq\pm1/2$.

Let us pause here and recall that we started from $\mathcal{H}_{XX}$, a model
of localized spins with nearest-neighbor coupling on a ring. The spinon spins,
by contrast, are no longer localized. Hence their interaction tends to be of
long range. For the situation under scrutiny here, all spinon spins are almost
equivalently coupled, somewhat reminiscent of electron spin couplings within
atomic shells.

Now we reduce the number $N_s$ of spinons in the $\pi/2$-orbital gradually. We
identify the fermion configurations of this $N_s$-spinon state for all possible
spinon-spin combinations. Then we calculate the reduced energy $\epsilon$ as a
function of $n_s\doteq N_s/N$ and $m_z$ in the limit of $N\to\infty$. The
resulting expression has the form
\begin{equation}
  \label{eq:29}
  \epsilon=\frac{1}{\pi}\cos(\pi m_z)
\left[2\sin\left(\frac{\pi}{2}\,n_s\right)-1\right]
\end{equation}
for $0\leq n_s\leq1$ and $|m_z|\leq n_s/2$, in generalization of (\ref{eq:21}).
The factor in square-brackets changes sign at $n_s=1/3$. The
equivalent-neighbor Ising model for the spinon-spin interaction is still
applicable. It is merely the effective coupling strength $J_e$ that now depends
on $n_s$ and that switches from ferromagnetic interaction at $n_s>1/3$ to
antiferromagnetic interaction at $n_s<1/3$, a Hund's rule of sorts. Inspection
of finite-$N$ $XX$ chains for which $n_s=1/3$ is realized indeed shows that in
those particular multiplets the spinon-spin coupling is absent. 

For a generic spinon orbital, specified by orbital momentum $\kappa$, the
fermion configuration consists of three clusters in momentum space:
\begin{equation}
  \label{eq:31}
  \left\{
\begin{tabular}{l}
 ${\displaystyle \frac{\pi}{2}-\pi m_z \leq p \leq \frac{3\pi}{2}
-\frac{\pi}{2}n_s-\pi
   m_z-\kappa}$,\\ \\
${\displaystyle \frac{3\pi}{2}+\frac{\pi}{2}n_s+\pi m_z
-\kappa \leq p \leq \frac{3\pi}{2}+\pi
  m_z}$,\\ \\
${\displaystyle \frac{3\pi}{2}-\frac{\pi}{2}n_s+\pi m_z 
+\kappa \leq p \leq \frac{3\pi}{2}+\frac{\pi}{2}n_s-\pi m_z+\kappa}$.
\end{tabular} \right\} 
\end{equation}
Integration of the fermion energy band, $\cos p$, over these three regions 
  yields
  \begin{equation}
    \label{eq:30}
    \epsilon =\frac{1}{\pi}\cos(\pi m_z)
\left[2\sin\left(\frac{\pi}{2}\,n_s\right)\sin\kappa-1\right]
  \end{equation}
  for $0\leq n_s\leq1$, $|m_z|\leq n_s/2$, and $\pi n_s/2\leq\kappa \leq\pi-\pi
  n_s/2$.  The effective equivalent-neighbor Ising exchange constant $J_e$ now
  depends on the filling $n_s$ and the orbital momentum $\kappa$.  A switch
  from ferromagnetic spinon spin coupling at low filling to antiferromagnetic
  coupling at higher filling exists only for orbitals with
  $\pi/4\leq\kappa\leq3\pi/4$. We have already seen that $J_e=0$ occurs at
  $n_s=1/3$ (amounting to capacity $n_s/n_s^\mathrm{max}=1/3$) for the orbital
  with $\kappa=\pi/2$. As we move away from that central orbital to either
  side, the value of $n_s$ at which $J_e=0$ occurs decreases gradually while
  the value of $n_s/n_s^\mathrm{max}$ increases. At $\kappa=\pi/4$ and
  $\kappa=3\pi/4$ we have $J_e=0$ at $n_s=1/2$, which means full capacity
  $(n_s/n_s^\mathrm{max}=1)$. The maximum capacity for fixed orbital momentum
  $0\leq\kappa\leq\pi$ is $n_s^\mathrm{max}= 1-|1-2\kappa/ \pi|$.

%
\subsection{Spinons in several orbitals}\label{sec:spintorb}
%
The most general spinon configuration involves $t$ orbitals with orbital
momenta $\nu_s\leq \kappa_1<\kappa_2<\cdots<\kappa_t\leq\pi-\nu_s$ and with the
spinon content of each orbital described by the variables
 \begin{equation}
    \label{eq:18}
    \nu_i\doteq\frac{\pi N_s^{(i)}}{2N},\quad 
\mu_i\doteq\frac{\pi M_z^{(i)}}{N},\quad
    i=1,2,\ldots,t;\quad \nu_s\doteq \sum_{i=1}^t\nu_i. 
  \end{equation}
The energy expression for this state can be rendered as follows:
\begin{eqnarray}
  \label{eq:Spinon20}
  && \hspace*{-20mm}
\pi\epsilon ~ = ~
  \sum_{i=1}^{t-1} \cos \bar{\mu}_{i,i+1}
  \left\{ \cos \bar{\nu}_{i,i+1} 
    \Big[ \cos\kappa_{i+1} - \cos\kappa_i \Big]
+\sin \bar{\nu}_{i,i+1}
    \Big[ \sin\kappa_i - \sin\kappa_{i+1} \Big]
  \right\}
  \nonumber \\ && \hspace*{-18mm}
  +\cos \bar{\mu}_{t,t+1}
  \left\{ \cos \bar{\nu}_{t,t+1}
    \Big[ \cos\kappa_1 - \cos\kappa_t \Big]
    +\sin \bar{\nu}_{t,t+1}
    \Big[ \sin\kappa_1 + \sin\kappa_t \Big]-1 \right\},
  \end{eqnarray}
where
\begin{equation}
  \label{eq:47}
  \bar{\mu}_{i,k}\doteq \sum_{j=1}^i \mu_j - \sum_{j=k}^t \mu_j,\qquad  
\bar{\nu}_{i,k}\doteq \sum_{j=1}^i \nu_j - \sum_{j=k}^t \nu_j.
\end{equation}
The first $t-1$ terms represent a coupling between nearest-neighbor spinon
orbitals in orbital momentum space. Each such term depends on the momenta of
the two coupled orbitals and on the ``local'' conserved quantities
$\bar{\mu}_{i,i+1}$, $\bar{\nu}_{i,i+1}$, which play the role of coupling
constants for nearest-neighbor orbitals. The last term has a slightly different
structure and depends on the smallest and largest orbital momentum values only.

%
\section{Thermodynamics of spinons}\label{sec:tdsp}
%
In this section we allow for the presence of a magnetic field $h$ in
$z$-direction as represented by terms $-hS_\ell^z$ added to the $XX$ Hamiltonian
(\ref{eq:17}) or by terms $-\mu c_{p_i}^\dagger c_{p_i}^{}$
added to (\ref{eq:12}). 
The thermodynamic properties of $\mathcal{H}_{XX}$ are derived with least
effort in the fermion representation \cite{Kats62}. From the grand partition
function of free fermions,
\begin{equation}
  \label{eq:4}
  Z=\prod_{p_i}\left(1+e^{-\beta(\cos p_i-\mu)}\right),\quad 
\beta=(k_{B}T)^{-1},
\end{equation}
we infer the grand potential per site in the limit $N\to\infty$,
\begin{equation}
  \label{eq:5}\hspace*{-15mm}
  \beta\omega(T,\mu)\doteq -\lim_{N\to\infty}N^{-1}\ln Z
=-\frac{1}{2\pi}\int_{-\pi}^{+\pi}dp\ln\left(1+e^{-\beta(\cos p-\mu)}\right), 
\end{equation}
which translates (for $\mu=-h$) into the Gibbs free energy per site in the spin
representation, $g(T,h)=\omega(T,\mu)+\mu n_f-hm_z$,
where $n_f\doteq N_f/N$ and $m_{z}=1/2-n_{f}$. This result is also obtainable
from magnons, namely via TBA applied to the $XXZ$ model in the limit
$\Delta\to0$ \cite{Taka99}.

Here we demonstrate a different thermodynamic analysis of $\mathcal{H}_{XX}$.
The results of Sec.~\ref{sec:basp} are the basis for an alternative TBA, not
from the top down via magnons but from the bottom up via spinons.
We introduce separate densities in momentum space for spinons with spin up and
spin down: $\rho_\sigma(k)$ for $\sigma=\pm$, respectively. For the total
number of spinons per site we write 
\begin{equation}
  \label{eq:14}
  n_s \doteq\sum_\sigma n_\sigma =  \frac{1}{2\pi} \sum_\sigma
\int_{k_\mathrm{min}^\sigma}^{k_\mathrm{max}^\sigma}
  dk\,\rho_\sigma(k),
\end{equation}
where $n_\sigma\doteq N_\sigma/N$. The integration limits are inferred from
(\ref{eq:2}):
\begin{equation}
  \label{eq:16}
  k_\mathrm{min}^\sigma=-\sigma\pi m_z,\qquad 
k_\mathrm{max}^\sigma=\pi(1+\sigma m_z).
\end{equation}
The magnetization $m_z$ (per site) is
itself expressible in terms of the $\rho_\sigma(k)$:
\begin{equation}
  \label{eq:24}
  m_z=\frac{1}{2}\sum_\sigma\sigma n_\sigma= \frac{1}{4\pi} \sum_\sigma
  \int_{k_\mathrm{min}^\sigma}^{k_\mathrm{max}^\sigma} dk\,\sigma\rho_\sigma(k).
\end{equation}
Given that all spinon momenta are distinct with allowed values equidistant on a
prescribed interval, as shown in Sec.~\ref{sec:basp}, we express the entropy
(per site) in the form
\begin{equation}
  \label{eq:32}\hspace*{-12mm}
  s=-\frac{k_B}{2\pi}\sum_\sigma 
\int_{k_\mathrm{min}^\sigma}^{k_\mathrm{max}^\sigma} dk
  \left[\rho_\sigma(k)\ln \rho_\sigma(k)
+\Big(1-\rho_\sigma(k)\Big)\ln\Big(1-\rho_\sigma(k)\Big)\right].
\end{equation}
The internal energy (per site) expressed via the $\rho_\sigma(k)$ follows
from (\ref{eq:49}):
\begin{equation}
  \label{eq:33}
  u= \frac{1}{2\pi}\sum_\sigma \int_{k_\mathrm{min}^\sigma}^{k_\mathrm{max}^\sigma}
  dk\,\rho_\sigma(k)\sin k -\frac{1}{\pi}\cos(\pi m_z).
\end{equation}
The spinon densities $\rho_\sigma(k)$ in thermal equilibrium minimize the Gibbs
free energy (per site), $g= u-Ts-hm_z$. Solving the variational problem,
$\delta g=0$, for the expression
\begin{eqnarray}
  \label{eq:11}
  && \fl g = \frac{1}{2\pi} \sum_\sigma 
\int_{-\sigma\pi m_z}^{\pi(1+\sigma m_z)}dk\Big\{ \rho_\sigma(k)\sin k
    +k_BT\Big[\rho_\sigma(k)\ln \rho_\sigma(k) \Big. \Big. \nonumber \\
&&  \Big. \Big. +\Big(1-\rho_\sigma(k)\Big)\ln\Big(1-\rho_\sigma(k)\Big)\Big] 
 -\frac{h}{2}\sigma\rho_\sigma(k)\Big\} -\frac{1}{\pi}\cos(\pi m_z),
\end{eqnarray}
assembled from (\ref{eq:14})--(\ref{eq:33}), is difficult unless we can remove
$m_z$ from the integration limits and from the argument in the last term.

A way out is suggested by the observation that the integration limits of
$\rho_+(k)$ and $\rho_-(k-\pi)$ are complements in the Brillouin zone
$[-\pi,\pi]$. Indeed, if we extend the domains (\ref{eq:16}) of both functions
$\rho_\sigma(k)$ to the full Brillouin zone via the relation
\begin{equation}
  \label{eq:25}
  \rho_+(k)+\rho_-(k-\pi)=1
\end{equation}
we can combine the two integrals in (\ref{eq:11}) into a single integral either
for $\sigma=+$ or $\sigma=-$ over the entire Brillouin zone with a slightly
modified integrand and the last term eliminated. Keeping the sum over $\sigma$
we write
\begin{eqnarray}
  \label{eq:34}
  g &=&  \frac{1}{4\pi}\sum_\sigma\int_{-\pi}^{+\pi}dk\Big\{ \rho_\sigma(k)\sin k
    +k_BT\Big[\rho_\sigma(k)\ln \rho_\sigma(k) \Big. \Big. \nonumber \\
&&  \hspace*{20mm}\Big. \Big. +\Big(1-\rho_\sigma(k)\Big)
\ln\Big(1-\rho_\sigma(k)\Big)\Big] 
 -h\sigma\rho_\sigma(k)\Big\}.
\end{eqnarray}
Now the extremum problem is readily solved. The spinon densities in thermal
equilibrium at temperature $T$ and magnetic field $h$ are
\begin{equation}
  \label{eq:35}
  \rho_\sigma(k)=\left[e^{\beta(\sin k-h\sigma)}+1\right]^{-1}.
\end{equation}
Substitution of (\ref{eq:35}) into (\ref{eq:34}) produces the explicit result
\begin{equation}
  \label{eq:77ad}
  \beta g(T,h)= -\frac{1}{2\pi}\int_{-\pi}^{+\pi}dk
\ln\left(2\cosh\Big(\frac{\beta}{2}(\sin k-h)\Big)\right), 
\end{equation}
consistent with the result (\ref{eq:5}) obtained via Jordan-Wigner
fermions. Hence the thermodynamics of the $XX$ model can be described entirely
via spinons.

%
\section{Conclusion and outlook}\label{sec:concl}
%

It is far from straightforward to generalize this study of the spinon
interaction and spinon thermodynamics to the $XXZ$ model at $\Delta\neq0$. The
results presented here for the $XX$ limit $(\Delta=0)$ set the stage for one
point from which to attack this challenge. A natural second point of attack is the
Ising limit $(\Delta=\infty)$, where the mapping between top-down
quasiparticles (ferromagnetic domains) and bottom-up quasiparticles
(antiferromagnetic domain walls) is again transparent and where the latter are
again spin-1/2 particles with semionic statistics \cite{LVP+07}.

A promising third point of attack is the Heisenberg limit $(\Delta=1)$. Its
spectral properties share key features with those of the HS model owing to
common symmetries. Given that the $XX$ and HS models, which have very different
symmetries, exhibit similar degrees of complexity regarding the quasiparticle
composition of their spectra from top down and from bottom up as well as
regarding the spinon interaction it is useful to compare several key results
established here for the $XX$ model with corresponding results known for the HS
model. Some relevant HS results are summarized in \ref{sec:appb}.
Noteworthy similarities and differences are pointed out along the way.


\appendix

%
\section{Haldane-Shastry model}\label{sec:appb}
\setcounter{section}{1}
%
The energy level spectrum of the HS model \cite{Hald88,Shas88},
\begin{equation}
  \label{eq:1hslj}
\mathcal{H}_{\rm HS} =\sum_{j<i}J_{ij}\mathbf{S}_i\cdot\mathbf{S}_j,\quad
  J_{ij}=J\left[\frac{N}{\pi}\sin\frac{\pi(i-j)}{N}\right]^{-2},
\end{equation}
on a ring of $N$ sites, is generated from the top down by so-called
pseudomomenta that represent Yangian multiplet states and satisfy the following
set of asymptotic BAE:
\begin{equation}
  \label{eq:3hslj}
  Np_i= 2\pi I_i +\pi\sum_{j\neq
      i}^M\mathrm{sgn}(p_i-p_j), \quad i=1,\ldots,M,
\end{equation}
where $0\leq M\leq [N/2]$ (integer part of $N/2$). The BQN $I_i$ are integers for
odd $M$ and half-integers for even $M$ on the interval
\begin{equation}
  \label{eq:4hslj}
  \frac{1}{2}(M+1)\leq I_i\leq N-\frac{1}{2}(M+1).
\end{equation}
The solutions of (\ref{eq:3hslj}) are of the form
\begin{equation}
  \label{eq:5hslj}
 p_i=\frac{2\pi}{N}\bar{m}_i,\quad \bar{m}_i\in \{1,2,\ldots,N-1\},\quad 
\bar{m}_{i+1}-\bar{m}_i\geq2.
\end{equation}
The wave numbers and energies of HS levels are
\begin{equation}
  \label{eq:8hslj}
k= \frac{2\pi}{N}\sum_{i=1}^M\bar{m}_i ~ ~\mathrm{mod}(2\pi),\quad
  E =2\left(\frac{\pi v_s}{N^2}\right)\sum_{i=1}^M\bar{m}_i(\bar{m}_i-N),
\end{equation}
where $v_s=\pi J/2$ and where the origin of the energy scale is set to coincide
with the vacuum of pseudomomenta $(M=0)$. The lowest energy level contains the
maximum number of pseudomomenta. This level is unique if $N$ is even and
fourfold degenerate if $N$ is odd. The pseudomomenta play a role similar to the
fermions in the $XX$ model.

From the bottom up the HS spectrum is generated by
spinons. The following relations hold between the number of spinons, $N_s$,
the number of pseudomomenta, $M$, and the magnetization, $M_z$:
\begin{equation}
  \label{eq:11hslj}
  N_+ +N_-= N_s=N-2M,\quad N_+ -N_-=2M_z.
\end{equation}
The number of spinons is determined by the number of pseudomomenta
alone. In $\mathcal{H}_{XX}$ the number of fermions alone does not determine the
number of spinons.  The unique HS ground state for even $N$ (spinon vacuum) has
energy
\begin{equation}
  \label{eq:51}
  E_\mathrm{sv}=-\left(\frac{\pi v_s}{N^2}\right)\frac{1}{6}N(N^2+2).
\end{equation}
The number of orbitals available for occupation by spinons is $(N-N_s)/2+1$,
where $N_s$ is even (odd) for even (odd) $N$.  The allowed
spinon orbital momenta are
\begin{equation}
  \label{eq:66hslj}
    \kappa_i=\frac{\pi}{N}m_i,\quad m_i=-M,-M+2,\ldots,+M.
\end{equation}

A generic HS eigenstate may be specified as follows:
\begin{equation}
\label{eq:55}
\left\{\begin{tabular}{c}
 $-\frac{1}{2}(N-N_s)\leq m_1\leq m_2\leq\cdots\leq m_{N_s}\leq
\frac{1}{2}(N-N_s)$ \\
 $\sigma_1,\sigma_2,\ldots,\sigma_{N_s},\quad \sigma_i=\pm$
\end{tabular}\right\}.
\end{equation}
The use of separate sets of quantum numbers for spinon momenta $(m_i)$ and
spinon spins $(\sigma_i)$ is the natural choice for the HS model. In the $XX$ model
we used separate sets of momentum quantum numbers for spin-up spinons and
spin-down spinons.  The wave number and the energy of the HS eigenstate
(\ref{eq:55}) are independent of the $\sigma_i$ and depend on the $m_i$ as follows
\cite{Hald91a,Tals95}:
\begin{equation}
  \label{eq:67hslj}
  k=M\pi +\frac{\pi}{N}\sum_{i =1}^{N_s}m_{i}~ ~\mathrm{mod}(2\pi),
\end{equation}
\begin{equation}
  \label{eq:12hslj}
E-E_\mathrm{sv}=E_{M}+\sum_{i=1}^{N_s}\epsilon(m_{i})
+\frac{1}{N}\sum_{i<j}V(m_{i}-m_{j}).
\end{equation}
The first term in (\ref{eq:12hslj}) depends (for given $N$)
only on the number of spinons present,
\begin{equation}
  \label{eq:13hslj}
  \hspace*{-15mm}E_{M}=\frac{\pi v_s}{N^2}\left\{\frac{1}{6}N(N^2+2)
    -\frac{1}{6}M\left[3N(N-1) -4M^2+6M+4\right]\right\},
\end{equation}
the second term depends on the orbital momenta,
\begin{equation}
  \label{eq:14hslj}
  \epsilon(m_i)=\frac{\pi v_s}{N^2}\left(M^2-m_i^2\right), 
\end{equation}
and the third term describes a pair interaction of sorts,
\begin{equation}
  \label{eq:15hslj}
 \frac{1}{N}V(m_i-m_j) =\frac{\pi v_s}{N^2}\left(M-|m_i-m_j|\right).
\end{equation}

A representation of Yangian multiplets that describes the pseudomomentum
content and the spinon content simultaneously is the \emph{motif} as
illustrated in Table~\ref{tab:motspinmag} for $N=5$. The motif of an HS
eigenstate consists of binary strings of length $N$. The elements of each
permissible string are a '10' (pseudomomentum) and a '0' (spinon).  All
consecutive '0's that do not belong to a '10' represent spinons in the same
momentum state. Consecutive '10's represent pseudomomenta with
$\Delta\bar{m}_i=2$. Every '0' between two '10's increases $\Delta\bar{m}_i$ by
one unit.  Pseudomomenta with increasing $\bar{m}_i$ are encoded by successive
'10's read from left to right. Spinons in orbitals with increasing $m_i$ are
encoded by successive '0's (separated by at least one '10') read from right to
left.

\begin{table}[htb]
  \caption{Motif, pseudomomentum quantum numbers $\bar{m}_i$,
    spinon orbital momentum quantum numbers $m_i$, wave number $k$ (in units of
    $2\pi/N$), 
    energy $E-E_\mathrm{sv}$ (in units of $\pi v_s/N^2$), spin content,
    degeneracy, and spinon momenta $k_i$ (in units of $\pi/N$) of all
    Yangian multiplets for $N=5$.}\label{tab:motspinmag} \vspace*{5mm}  
\centerline{\small
\begin{tabular}{c|cc|cc|c|c|c}
motif & $\bar{m}_i$ & $m_i$ & $k$ & $E-E_\mathrm{sv}$ &
spin & deg. & $k_i$ \\ \hline
$0\,10\,10$ & $2,4$ & $2$ & $1$ & $\frac{5}{2}$ &
$\frac{1}{2}$ & $2$ & $2$ \rule[-2mm]{0mm}{6mm} \\
$10\,0\,10$ & $1,4$ & $0$ & $0$ & $\frac{13}{2}$ &
$\frac{1}{2}$ & $2$ & $0$ \rule[-2mm]{0mm}{5mm} \\
$10\,10\,0$ & $1,3$ & $-2$ & $4$ & $\frac{5}{2}$ &
$\frac{1}{2}$ & $2$ & $-2$ \rule[-2mm]{0mm}{5mm} \\
$00\,10\,0$ & $3$ & $-1,1,1$ & $3$ & $\frac{21}{2}$ &
$\frac{1}{2}\otimes1=\frac{3}{2}\oplus\frac{1}{2}$ & $6$ & $-3,1,3$ 
\rule[-2mm]{0mm}{5mm} \\
$0\,10\,00$ & $2$ & $-1,-1,1$ & $2$ & $\frac{21}{2}$ &
$\frac{1}{2}\otimes1=\frac{3}{2}\oplus\frac{1}{2}$ & $6$ & $-3,-1,3$ 
\rule[-2mm]{0mm}{5mm} \\
$000\,10$ & $4$ & $1,1,1$ & $4$ & $\frac{29}{2}$ &
$\frac{3}{2}$ & $4$ & $-1,1,3$ \rule[-2mm]{0mm}{5mm} \\
$10\,000$ & $1$ & $-1,-1,-1$ & $1$ & $\frac{29}{2}$ &
$\frac{3}{2}$ & $4$ & $-3,-1,1$ \rule[-2mm]{0mm}{5mm} \\
$00000$ & -- & $0,0,0,0,0$ & $0$ & $\frac{45}{2}$ &
$\frac{5}{2}$ & $6$ & $-4,-2,0,2,4$ \rule[-2mm]{0mm}{5mm} 
\end{tabular}
}
\end{table} 

The spin content of any given Yangian multiplet can be read off the binary
motif by recognizing the multiplets of the quantum number $S_T$ representing
the total spin in each spinon orbital. In the $XX$ case the motif pertains to
individual eigenstates (Fig.~\ref{fig:fermspinN5}) and encodes a specific
spinon spin configuration. 

Just as in the $XX$ model, the energy expression (\ref{eq:12hslj}) rewritten in
the form 
 \begin{equation}
    \label{eq:52}
    E-E_\mathrm{sv}= E_0(N_s) - 
\frac{\pi v_s}{N^2}\sum_{i=1}^{N_s}m_{i}(m_{i}-N_s-1+2i),
  \end{equation}
  \begin{equation}
    \label{eq:53}
    E_0(N_s)=\left(\frac{\pi v_s}{N^2}\right)
\frac{1}{12}N_s\left(3N^2+4-N_s^2\right).
  \end{equation}
  is suggestive of a CBA for spinons.  The $m_i$ with range (\ref{eq:55})
  become the BQN and (\ref{eq:52}) turns into
\begin{equation}
  \label{eq:56}
  E-E_\mathrm{sv}= E_0(N_+ +N_-) -\frac{v_s}{\pi}\sum_{j=1}^{N_s}\kappa_{j} k_{j},
\end{equation}
where the spinon momenta 
\begin{equation}
  \label{eq:62}
  k_i=\frac{\pi}{N}\left(m_{i}-N_s-1+2i\right)
\end{equation}
are the solutions of the BAE:
\begin{equation}
  \label{eq:57}\hspace*{-5mm}
  Nk_i= \pi m_i +\sum_{j=1}^{N_{s}}\theta_{HS}(k_i-k_j),\quad
  i=1,\ldots,N_s,
\end{equation}
\begin{equation}
  \label{eq:58}\hspace*{-5mm}
  \theta_{HS}(k_i-k_j)=\pi\,\mathrm{sgn}(k_i-k_j).
\end{equation}
Two differences from the $XX$ model, Eqs.~(\ref{eq:50})-(\ref{eq:46}), are
noteworthy: (i) we are dealing with just one set of BQN; (ii) all spinons
scatter off each other, not just those with the same spin orientation.

The HS model too can be interpreted as a set of spinon orbitals, each specified
by an orbital momentum $\kappa_i$. Each such orbital is filled with spinons of
arbitrary spin orientation up to a certain capacity. The available orbitals
again depend on the total number of spinons present.  The energy expression for
any Yangian multiplet only depends on the orbital momenta and fillings.
Conclusions can again be drawn about the energetics and interaction of spinon
orbitals.

Beginning with the case where all spinons are in the same orbital, we express
the reduced energy, $\epsilon\doteq (E-E_\mathrm{sv})/(\pi v_sN)$, as a
function of $n_s\doteq N_s/N$ and $\bar{\kappa}\doteq\kappa/2\pi$ in the limit
$N\to\infty$:
\begin{equation}
  \label{eq:54}
  \epsilon=\frac{1}{12}n_s\left(3-n_s^2-48\bar{\kappa}^2\right),
\end{equation}
for comparison with the $XX$ result (\ref{eq:30}), which also depends on the
magnetization.
The generalization of the HS expression (\ref{eq:54}) to $t$ spinon orbitals is
again structurally similar to the $XX$ result
(\ref{eq:Spinon20}), except for the absent spinon-spin dependence:
\begin{eqnarray}
  \label{eq:HS20}
  && \fl \epsilon  =   \sum_{i=1}^{t-1} 
  \frac{8}{3}\left[ (a_{i,i+1}-\bar{\kappa}_i)^3
-(a_{i,i+1}-\bar{\kappa}_{i+1})^3 \right]
  +2\left[ (a_{i,i+1}-\bar{\kappa}_{i+1})^2-(a_{i,i+1}-\bar{\kappa}_i)^2 \right]
  \nonumber \\
  &&\hspace*{-15mm}
  +\frac{8}{3}\left[ (a_{t,t+1}-\bar{\kappa}_t)^3-(a_{0,1}-\bar{\kappa}_1)^3 \right]
  +2\left[ (a_{0,1}-\bar{\kappa}_1)^2-(a_{t,t+1}-\bar{\kappa}_t)^2 \right],
\end{eqnarray}
where $\nu_j\doteq N_s^{(j)}/(2N)$, 
\begin{equation}
  \label{eq:HS12}
  a_{i,k} \doteq \frac{1}{4} - \frac{1}{2}\sum_{j=1}^i \nu_j 
  + \frac{1}{2}\sum_{j=k}^{t} \nu_j, \quad i=0,1,2,\ldots,t,
\end{equation}
and $N_s^{(j)}$ is the number of spinons (with arbitrary spin orientation) in
the orbital with reduced momentum $\bar{\kappa}_j$.

The thermodynamics of spinons for $\mathcal{H}_\mathrm{HS}$ was reported by
Haldane along two different paths. The approach taken in Ref.~\cite{Hald91}
uses the spinon orbital momenta $\kappa_i$ (named $k$) as the independent
variables. Since there are no restrictions on the occupation of available
spinon orbitals for given $N_s$, a bosonic version of the entropy functional is
used.  The approach analogous to the one taken in Sec.~\ref{sec:tdsp} for
$\mathcal{H}_{XX}$ would use the spinon momenta $k_i$ as the independent
variables. Since all $k_i$ are distinct, a fermionic version of the the entropy
functional would have to be used as in (\ref{eq:32}). The approach taken in
Ref.~\cite{Hald94} introduces rapidities, $x_i=(2m_i+2i-N_s)/N$, which again
are all distinct, but have the advantage of being confined to an
interval with limits that are independent of $N_s$.
Haldane's result for the Gibbs free energy per site,
  \begin{equation}
    \label{eq:65ad}
    \beta g=-\frac{1}{2}\int_{-1}^{+1}dx\,
\ln\left(\frac{\sinh\left(\frac{\beta h}{2}[1+\mu(x)]\right)} 
{\sinh\frac{\beta h}{2}}\right), 
  \end{equation}
\begin{equation}
    \label{eq:66ad}
    \sinh\left(\frac{\beta h}{2}\mu(x)\right)
=\sinh\frac{\beta h}{2}e^{-\beta\epsilon_{0}(x)},\quad 
    \epsilon_{0}(x)=\frac{1}{4}\pi v_s(1-x^2).
  \end{equation}
is to be compared with the corresponding $XX$ result (\ref{eq:77ad}). The
distributions of spinon rapidities inferred from (\ref{eq:65ad}), 
\begin{equation}
  \label{eq:75ad}
  n_\sigma(x)=\frac{\sigma}{2}
\frac{\tanh\left(\frac{\beta h}{2}\mu(x)\right)}{1-e^{-\sigma 
\beta h[1+\mu(x)]}},\quad
|x|\leq1, 
\end{equation}
are to be compared with the distribution of spinon momenta (\ref{eq:35}) in
the $XX$ model. 

\vspace*{5mm}
%
\ack
%
Financial support from the DFG Schwerpunkt \textit{Kollektive Quantenzust{\"a}nde
  in elektronischen 1D {\"U}bergangsmetallverbindungen} (for M.K.)  and from the Graduiertenkolleg
Darstellungstheorie und ihre Anwendungen in Mathematik und Physik (for KW) is gratefully
acknowledged.

\vspace*{5mm}




\end{document}